# Geometry-covering Jammer Localization based on Distance Comprehension in Wireless Sensor Networks

*Sheng. Wang*
Tianjin University
Tianjin, China
shengtju@foxmail.com

*Chunliang. Chu*
Tianjin University
Tianjin, China
ccl_tianda@139.com

*Abstract*—Jamming attacks could cause severe damage to Wireless Sensor Networks (WSNs). Once jamming attack occurs, the most urgent work is to get the position information of the jammer. Then safety measures to eliminate the jamming effects can be devised. In this paper, the jammer localization is conducted by geometric covering method to achieve a low energy consumption. And utilizing the power of the jamming signal received by the boundary nodes, a compensating method is composed to reduce the estimating error of the jamming area. At last the localization is conducted by extracting the minimum covering circle of the compensated victim area. Simulations are conducted to test the localization accuracy with the impact of node density, jamming region and radius. Results show that this localization method achieves both good precision and low energy consumption.

*Keywords*—*wireless sensor networks, jamming attack, localization, network safety*

## I. INTRODUCTION

Wireless networks are becoming increasingly pervasive. Many devices such as Wi-Fi network adapters, infrared remote device, Bluetooth headset and microwave ovens in our daily lives utilize wireless technology. Wireless technology is changing the way we communicate and intersect with the world. For example, we can monitor wild animals remotely and track the positions of different targets employing wireless technology[1]. However, The wireless communication is based on the share of privacy[2], which makes the Wireless Sensor Networks (WSNs) easy to be attacked. As a result, jamming attacks often occur and interfere the messages from transmitting. The wireless network is blocked from sending out messages. Unpredictable wireless devices and enemies often cause jamming among nodes and interfere wireless broadcasting.

To improve the reliability of the WSN, safety measures need to be taken to ease the jamming effects. At the physical layer, several technologies have been proposed to mitigate the damage, such as DSSS(Direct Sequence Spread Spectrum)[3] and FHSS(Frequency Hopping Spread Spectrum)[4]. However, due to the energy limit in wireless sensor networks, these technologies are not suitable for WSNs because of too much energy cost. Specific defending method for the interfere need to be proposed. After jamming attack occurs, the most urgent work is to get the position information of the jammer. Then further security measures can be taken to eliminate the jamming effect. Such as defining a specific routing protocol that can choose a transmitting path to evade the jamming area[5] or removing the jammer from the network physically.

In WSNs, localizing the position of the jammer is not the same as the sensor nodes' localizing problem. Most localization technologies require the wireless sensor nodes equipped with unique hardware, such as ultrasound and infrared equipment. Other localizing methods need the information of neighbor nodes, but the nodes near the jammer cannot send out messages. The neighbor nodes will not cooperate. So the localization need to be conducted without the information of jammed nodes. Considering the specific localization situation, the convex hull study in geometry draws attention. The minimum covering circle method was applied to localized the jammer. They tried to find the minimum circle covering the boundary nodes[6]. The center of the covering circle is the estimated position of the jammer. This method showed lower algorithm complexity but cannot achieve high accuracy.

In this paper, we proposed a geometric-covering method which can be compensated by the jamming strength. The disadvantage of the existing geometric method is the complete dependence on the boundary nodes. Actually, the boundary nodes are not always located on the edge of the jamming area. So the real jamming area should be a larger area than the circle estimated by the boundary nodes. In our localization algorithm, the jamming strengths on the boundary nodes are taken into consideration. The covering area is extended towards the real jamming area. The minimum covering circle is conducted by the two maximum distance between boundary nodes.

Simulation shows that our geometric covering method algorithm performs well and achieve both high accuracy and low energy consumption.

The reminder of this paper is as follows. The related work is discussed in Section II. In Section III, we introduce the wireless network and the jamming model first. Then the analysis and description of our algorithm are given. In Section IV, we conduct simulations and analyze the impact of different factors(node density, jamming region and radius) on the localization. In the end, we conclude in Section V.

## II. RELATED WORKS

The WSN is gaining its popularity in many aspects. The mobile phone has become an essential device for modern life as a wide use of wireless network. In many occasions, the position information is needed by the network appliance. So the wireless localization problem arise experts' attention. Many algorithms has been proposed to obtain the position information of a device. In the field of wireless network localization, algorithms can be classified by the use or not of extra equipments. Hopper et has conducted the localization with infrared equipments in [7]. Ultrasound infrastructure has been analyzed as well by Ward et in [8]. Both of these two algorithms need the help of specific infrastructure to achieve the localization. While the localization method based on the received signal strength(RSS) could achieve the localization without the use of the extra equipment[9]. Besides, the localization algorithms can be categorized into range-based and range-free. The former method estimate the distance between nodes by physical technologies such as RSS. While range-free methods[10] deploy coarser metrics to estimate the position.

In the area of jammer localization, there are some algorithms already proposed. Pelechrinis et proposed to carry out the localization by performing gradient descent search in [11], but they didn't present the performance evaluation. Liu proposed to localize the jammer utilizing the network topology changes and introduced the concept of virtual forces to estimate the jammer's position in [12]. The virtual forces are defined by the states of sensor network and guide the estimated location of the jammer towards its accurate position iteratively. Liu conducted the localization utilizing the change of the nodes' hearing ranges caused by the jamming attack in [13]. He also proposed a error minimizing framework to improve the localization accuracy. These algorithms all depend on highly complex computation, causing huge energy waste. Sun proposed to localize the jammer by seeking for the minimum covering circle in [14]. The existing geometric method cannot localize jammer accurately because it did not take the jamming strength into accounts but only deployed the positions of the boundary nodes. Actually, the boundary nodes are not always located on the edge of the jamming area. So the existing geometric method cannot achieve high accuracy. Our localization algorithm is different from the existing algorithms and achieves both good precision and low energy consumption.

## III. PROBLEM FORMATION

Considering the various effects jammer may cause to the wireless network. It's impractical to come up with a universal model which fits for all the jamming occasions. The jammer constantly broadcast useless messages at a unique power level, forming a circle area in which the sensor nodes are blocked. While on the edge of the jamming area, the nodes could communicate with its neighbor nodes due to the lower jamming power.

Besides, the network is able to detect the jamming. The network can identify the jamming attack and obtain the network situation. The jammer, has an omnidirectional antenna. So the jamming ranges in all direction are the same and the jamming power will not change dramatically. In this paper, the localization algorithm only deal with single jammer or multiple jammers that don't have overlapping jammed area. Before we present our localization algorithm, we shall explain our network model clearly. The sensor network model is as follows.

### A. Network Model

Consider the wireless network in use, such as the Mica-2 sensor networks. As the initial work, the jammer localization are proposed for specific network model with the following characteristics.

I    Stationary. Once the jamming occurred, the location of each nodes remains unchanged. At least they remain stay until the network topology information is collected.

II    Location-aware. The coordinate of each sensor nodes can be obtained if it is not jammed. This could be realized in most sensor networks as most applications require location service.

III    Power-sensitive. Sensor nodes have the ability to perceive the power message send by the other nodes and send the power information out to its neighbor nodes.

Our research focus on the localization of the jammer, the scheme how the jamming attack is detected is not discussed. The detection method can be referred in [15].

### B. Jamming Model

In fact, most jammer have isotropic effect. The jammed region is a circular area centered at the jammer. The nodes are affected variously due to their different positions in the network. Each nodes received specific disturbance. And the dividing line is the edge of the jamming effects. According to the different impact they received, the sensor network nodes can be classified into three categories:

Unaffected Node. These sensor nodes are outside the jammed area. The jammer doesn't have influence on them. . These nodes can communicate with all their neighbors without any changes,and can function normally.

$N_u = \{n_u | d_{Ji} > R_J, \forall i \in n\}$

Boundary Node. The boundary nodes suffer from the jamming attack, but they are near to the edge of the jamming area. Their neighbor list is changed, but the condition for their broadcasting is still satisfied. So they can communicate with their unaffected neighbor nodes. In our algorithm, the boundary nodes play a important role. They are used to construct the convex hull and conduct the minimum covering circle. $N_B = \{n_b | R_J - r_n \leq d_{Ji} \leq R_J, \forall i \in n\}$

Jammed Node. The jammed nodes are inside the jamming area. They are totally blocked by the jammer due to the disturbance of the jamming signal. As a result, this nodes cannot receive and send out information from the majority of the networks. In other words, this nodes have been eliminated in the topology of the jammed network.

$N_J = \{n_J | d_{Ji} < R_J - r_n, \forall i \in n\}$

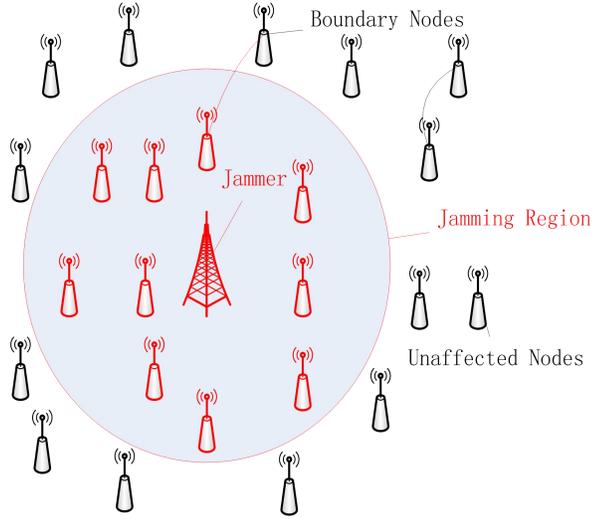

Fig. 1. A simple model of jammed network.

**Table 1.** Description of variables.

| Symbol | Description of variables |
|---|---|
| $P_{r_i}$ | Signal strength at boundary node i |
| $P_{J_j}$ | Transmission power of jammer j |
| K | Unit-less constant defined by the attenuation characteristics |
| $(X_J, Y_J)$ | Coordinates of the jammer |
| $(X_i, Y_i)$ | Coordinates of a boundary node i |
| $d_{Ji}$ | The distance between the jammer and the node i |
| $d_{ij}$ | The distance between the node i and node j |
| $R_J$ | The jamming radius of the jammer |
| $n_i$ | The node i |

*C. Problem analysis*

In wireless broadcasting, the signal strength decreases as the distance increases due to the path loss and shadowing. Path loss have a close relationship with the broadcasting distance. As for the jamming broadcasting, the path loss and shadowing remain exists. With a larger distance from the jammer, the signal strength become more weakened.

The network broadcasting model are built referring to [16] and [17]. The widely used log-normal shadowing model capture the essential of the path loss and shadowing. The model definition is as follows. Let $P_f$ be the received signal power subject to path loss attenuation alone and the strength of a transmitted signal be $P_t$. The received signal power at a distance of d can be modeled as the sum of $P_f$ and a variance (denoted by $X_\sigma$) caused by shadowing and other random attenuation.

$$P_r = P_f + X_\sigma \quad (1)$$
$$P_f = P_t + K - 10\eta \log_{10}(d) \quad (2)$$

$X_\sigma$ is a Gaussian zero-mean random variable with standard deviation σ, K is a unit-less constant depended on the antenna characteristics, η is the Path Loss Exponent. In a free space, η is 2 and $X_\sigma$ is always zero.

Once the jamming occurs, the wireless network topology changes and the message that jamming attack happens are sensed. A new network topology information will be obtained(shown in Fig. 2.). The algorithm need to employ the acquired information to localize the jammer. The minimum covering circle method is involved. In geometry, we know that the center of a circle must locate on the perpendicular bisector of the segment obtained by any two points on the circle. In the jammed network, the jammed region is a circular area centered at the jammer's position.

The Centroid Localization[18] (CL) algorithm uses the positions of all boundary nodes to locate the jammer. For example, if there are n nodes{$(X_1, Y_1), (X_2, Y_2), \cdots\cdots (X_N, Y_N)$}, the position will be estimated at$(X_J, Y_J)$:

$$(X_J, Y_J) = (\frac{\sum_{i=1}^{N} X_i}{N}, \frac{\sum_{i=1}^{N} Y_i}{N}) \quad (3)$$

Consider the random distribution of the boundary nodes, the Centroid Localization may not achieve precise positioning in some cases.

The Catch the Jammer(CJ) algorithm proposes to find the minimum covering circle of the jammed area[14]. A local convex hull formed by the boundary nodes is constructed. The center of the circle is the estimate position of the jammer. However, some boundary nodes are not on the edge of the jammed area. There exist some deviation in determining the jamming area. Utilizing the position of this nodes alone to locate the jammer cause localization error.

Consider the jamming strength on the boundary nodes, the estimated circle area can be extended towards the real jamming area. In this way, our convex hull can appropriately cover the jamming area, and achieves higher accuracy.

*D. Algorithm description*

The jamming localization algorithm Geometrical Jammer Localization(GJL) deploys the boundary nodes of the jamming network. Consider the jamming strength on the boundary nodes, we compensate the estimated effecting area of the jammer. The following is our localization process.

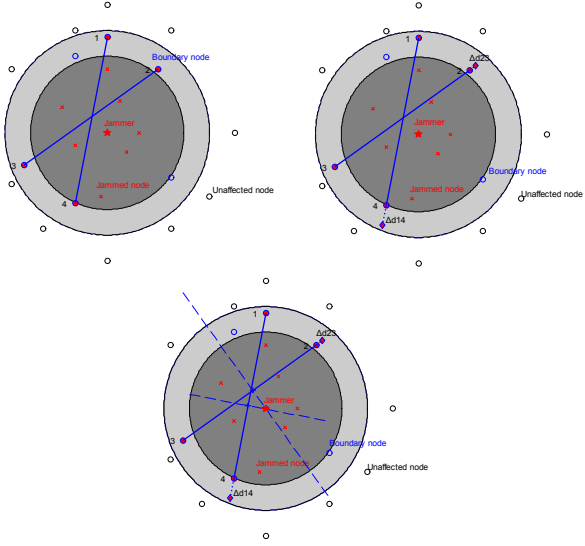

Fig.2(a). The Schematic Diagram of Distance Compensation and Jammer Localization.

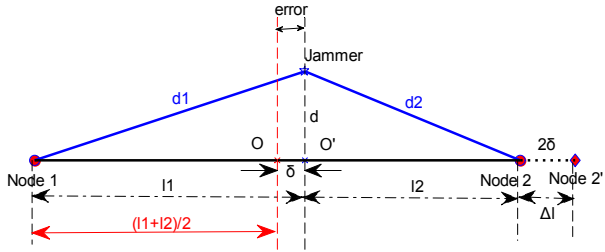

Fig.2(b). The Positions of the Jammer and Boundary Nodes.

Once jamming occurs, the first step we need to take is to employ the network. Get the information of the boundary nodes and the corresponding sensed power. A topology network as Fig. 2(a). can be obtained. The position information of all the boundary nodes and unaffected nodes can be utilized for the localization.

Then the distances between each two boundary nodes are calculated and compared. The two pairs of boundary nodes that have the maximum distance can be acquired. The two line segments are most likely to pass the center of the circle, shown in Fig. 2(b).

Consider the jamming power on the boundary nodes, the two line segments should be extended. Refer to the broadcasting characteristic in the network, we should follow appropriate rules. According to the log-normal shadowing model, the power received by nodes can be defined as follows.

$$P_1 = P_J + K - 10\eta \log_{10}(d_{J1}) \quad (4)$$
$$P_2 = P_J + K - 10\eta \log_{10}(d_{J2}) \quad (5)$$

Equation(4) minus Equation(5), $\eta = 2$, we can obtain

$$P_2 - P_1 = 10\eta \log_{10}(\frac{d_{J1}}{d_{J2}}) \quad (6)$$

$$P_2 - P_1 = 10\eta * \log_{10}(\frac{\sqrt{l_1^2 + d^2}}{\sqrt{l_2^2 + d^2}}) \quad (7)$$

After simplification

$$\Delta l = l_1 - l_2 = \frac{kd_{12}}{2+k}, k = 10^{\frac{P_2 - P_1}{20}} \quad (8)$$

The value of d12 can be obtained. $\Delta l$ is the length that should be extended to the line segments.

At last, the localization is conducted by the two segments. The two perpendicular bisector of the lines are drawn and the estimating jammer position is on the joint.

## IV. EXPERIMENTAL RESULTS

In this section, the performance of GJL algorithm is evaluated. The wireless sensor network area is a square field of 100meter by 100meter. The sensor nodes are randomly distributed in the area. Comparing with CJ and CL, simulations are conducted under different network situations (nodes' density, jamming region and jamming radius). We take the localization error ($\Delta$)as the simulation result. It is the distance between the estimated location of the algorithm $(\overline{X}_J, \overline{Y}_J)$ and the real position of the jammer $(X_J, Y_J)$. The error is defined as:

$$\Delta = \sqrt{(X_J - \overline{X}_J)^2 + (Y_J - \overline{Y}_J)^2} \quad (9)$$

### A. Impact of the Node Density

Node density plays a important role in jammer localization. {50, 100, 150, 200}nodes are distributed randomly in a square field of 100meter by 100meter. The jammer is not fixed but distributed randomly around the center of the field. The jamming radius are fixed to 30meter and the number of nodes in the network is 100.

The following Fig. 3. shows the localization error of different algorithm. The error decreases as the nodes' number increase. As the node density increases, the number of the boundary nodes increases. Causing more position information can be utilized in the localization algorithm. However, due to the randomly distribution, CL shows its instability. The same unstable results are given by CJ. The error simulated by our algorithm is more stable, and outperforms the other algorithms, reduced the localization error of the Centroid Localization by 70% on average.

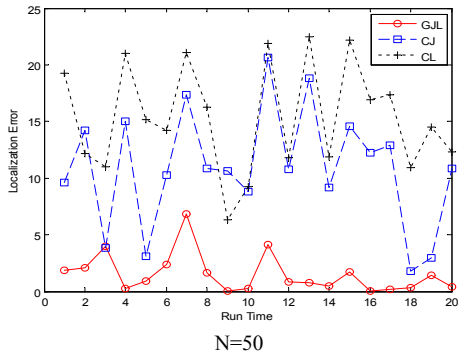

N=50

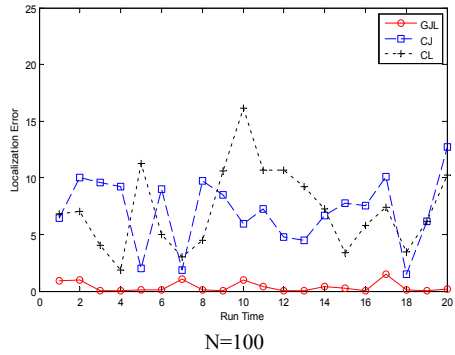

N=100

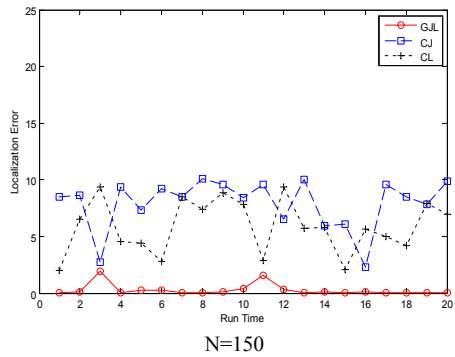

N=150

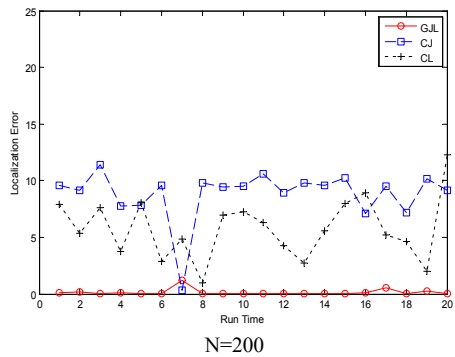

N=200

Fig.3. Impact of the node density on localization.

*B. Impact of the Jamming Region*

The jammer can be distributed in different regions of the network. Taking the Edge Effects into consideration, algorithms may be sensitive to the position of the jammer. In our simulation, we simulate the jammer in different region of the network(shown in Fig. 4). The jamming radius is fixed at 30meter. The performance of the algorithms is shown in Fig. 5.

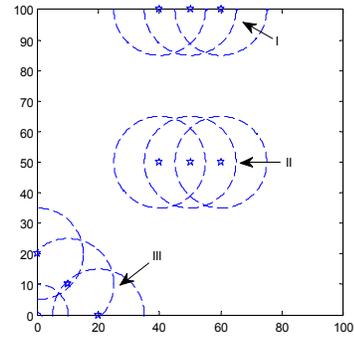

Fig.4. Different Distribution of Jamming Region.

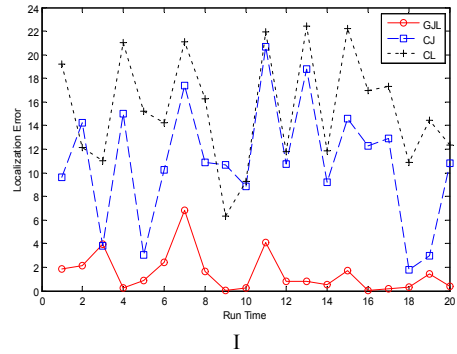

I

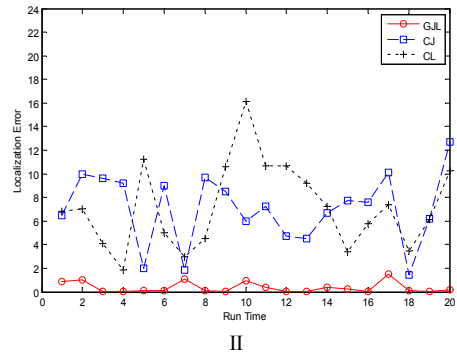

II

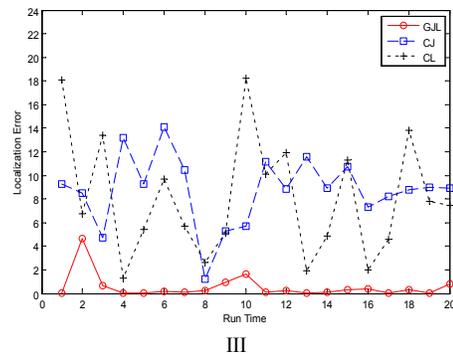

III

Fig.5. Impact of the Jamming Region on Localization error.

When the jammer is on the edge of the network, despite the symmetrical effect, the jamming area is not a circle that can be searched. The boundary nodes cannot construct a complete jamming circle. CJ cannot achieve accurate localization. In the simulation, all the algorithms show instability in this situation. The estimation error of the jamming area lend to the localization error. While taking the jamming strength into consideration, our algorithm shows the best accuracy.

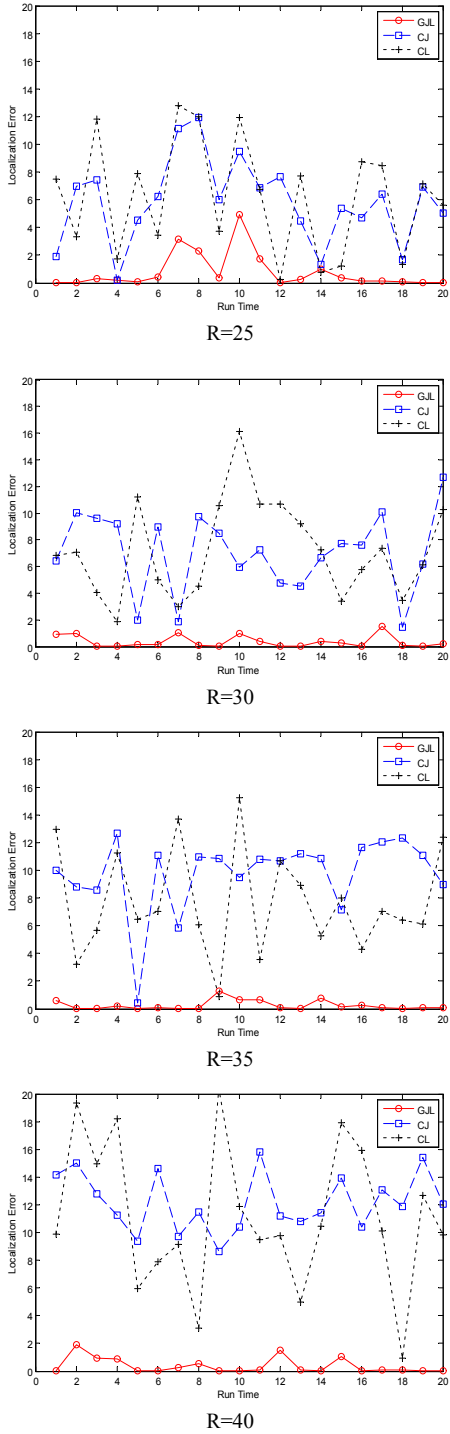

Fig.6. Impact of Jamming Radius on Localization error

*C. Impact of the Jamming Radius*

The jamming power is another affecting factor, which decides the size of the jamming area. In the simulation, the performance of GJL at different jamming radius{20 30 40m} are evaluated. In fact, the increase in jamming radius is the same as the increase in node density. Due to the larger number of boundary nodes involved in the localization, all the algorithms show higher accuracy. And still, it can be proved in the simulation that GJL outperforms the other algorithms.

## V. CONCLUSION

In this paper, a new jammer localization algorithm GJL is proposed. It is one kind of geometric method, combined with the compensation of jamming strength. The jamming area is estimated accurately. The method shows lower computation complexity and achieves good energy efficiency. In section IV, simulations of the localization algorithm are presented. With the simulation under different situations of node density, jamming region and radius, the universal applicability of GJL is proved. GJL achieves better localization precision at a lower energy consumption.

Our work explored the one jammer localization scenario. The problem of directional jammer and multiple jammer localization would be our ongoing work. We also plan to exploit the feasibility to adapt our scheme into more sophisticated jamming attack models.